\pdfoutput=1
\documentclass[aps,prb,twocolumn,amsmath,amssymb,superscriptaddress]{revtex4-1}
\usepackage[T1]{fontenc}
\usepackage[latin9]{inputenc}
\usepackage[english]{babel}
\usepackage{graphicx}
\usepackage[caption=false]{subfig}
\usepackage{amssymb}
\usepackage{color}
\usepackage{bibunits}
\usepackage[hyperindex,breaklinks]{hyperref}
\usepackage{bm}

\begin{document}

\title{Resolving the VO$_2$ controversy: Mott mechanism dominates the insulator-to-metal transition}

\author{O. N\'ajera}
\affiliation{Laboratoire de Physique des Solides, CNRS-UMR8502, Universit\'e
Paris-Sud, Orsay 91405, France}

\author{M. Civelli}
\affiliation{Laboratoire de Physique des Solides, CNRS-UMR8502, Universit\'e
Paris-Sud, Orsay 91405, France}

\author{V. Dobrosavljevi\'{c}}
\affiliation{Department of Physics and National High Magnetic Field Laboratory,
Florida State University, Tallahassee, FL 32306, USA}

\author{M. J. Rozenberg}
\affiliation{Laboratoire de Physique des Solides, CNRS-UMR8502, Universit\'e
Paris-Sud, Orsay 91405, France}

\pacs{71.27.+a 71.30.+h, 78.20.-e}
\begin{abstract}
We consider a minimal model to investigate the metal-insulator
transition in VO$_2$. We adopt a Hubbard model with two orbital per
unit cell, which captures the competition between Mott and
singlet-dimer localization. We solve the model within Dynamical Mean
Field Theory, characterizing in detail the metal-insulator transition
and finding new features in the electronic states. We compare our
results with available experimental data obtaining good agreement in
the relevant model parameter range. Crucially, we can account for
puzzling optical conductivity data obtained within the hysteresis
region, which we associate to a novel metallic state characterized by
a split heavy quasiparticle band. Our results show that the
thermal-driven insulator-to-metal transition in VO$_2$ is compatible
with a Mott electronic mechanism, providing fresh insight to a long
standing ``chicken-and-egg'' debate and calling for further research
of ``Mottronics'' applications of this system.
\end{abstract}

\maketitle

\begin{bibunit}[unsrtnat]
Vanadium dioxide VO$_2$ and vanadium sesquioxide V$_2$O$_3$ remain at
the center stage of condensed matter physics as they are prototypical
examples of systems undergoing a strongly correlated metal-insulator
transition (MIT) \cite{Imada1998}. Their unusual electronic behavior
makes them very attractive materials for novel electronic devices
\cite{ahn_electrostatic_2006,Takagi2011}. In fact, they are
intensively investigated in the emerging field of ``Mottronics'', which
aims to exploit the functionalities associated to the quantum Mott
transitions. A key goal is to create fast and ultra-low power
consumption transistors, which may be downsized to the atomic limit
\cite{VO2-FET,Nakano2012,Jeong2013}.

VO$_2$ and V$_2$O$_3$ have nominally partially filled bands, hence are
expected to be metals. However, they undergo a first order metal to
insulator transition upon cooling at $\sim$ 340K and 180K,
respectively.  This phenomenon has been often associated to a Mott MIT
\cite{Imada1998}, namely a transition driven by the competition
between kinetic energy and Coulomb repulsion \cite{Mott}.  Yet, that
point of view has been questioned as
often\cite{Imada1998,Hansmann2013}.

The case of VO$_2$, displaying a transition from a high-$T$ rutile (R)
metal to a low-$T$ monoclinic (M$_1$) insulator, is emblematic
\cite{Morin1959, Goodenough1960, Bongers1965, Qazilbash2007,
Wegkamp2014, Aetukuri2013, Arcangeletti2007, Laverock2014}. The
central issue is whether the transition is driven by a spin-Peierls
structural instability, or by the electronic charge localization of
the Mott-Hubbard type.  This issue has been under scrutiny using
electronic structure calculations \cite{Biermann2005, Weber2012,
Brito2015, Silke2008, Silke2009} based on the combination of density-functional theory
in the local-density approximation with dynamical mean-
field theory (LDA+DMFT)
\cite{KotliarRMP2006}.  In the pioneering work of ref.
\cite{Biermann2005}, Biermann et al. argued that the insulator should
be considered as a renormalized Peierls insulator. Namely, a
band-insulator where the opening of the bonding-antibonding gap is
driven by dimerization and renormalized down by interactions
\cite{Biermann2005}.  On the other hand, the calculations showed that
within the metallic rutile phase, the Coulomb interaction failed to
produce a MIT for reasonable values of the interaction. More recently,
the problem was reconsidered by Brito et al. \cite{Brito2015} and by
Biermann et al. as well \cite{Silke2008, Silke1, Silke2, Silke3}
providing a rather different scenario.  Brito et al. found a MIT
within a second monoclinic (M$_2$) phase of VO$_2$ that only has half
the dimerization of the standard M$_1$, for the same value of the
Coulomb interaction.  Hence, they argued that Mott localization must
play the leading role in both MITs.  Nevertheless, they also noted
that the Mott insulator adiabatically connects to the singlet dimer
insulator state, and therefore the transition should be considered as
a Mott-Hubbard in the presence of strong inter-site exchange
\cite{Brito2015, Silke2008, Silke1}.

While those LDA+DMFT works provided multiple useful insights, the
issue whether the first order MIT at 340K in VO$_2$ is electronically
or structurally driven, still remains.  Here we shall try to shed new
light on this classic "chicken-and-egg" problem by adopting a
different strategy.  We shall trade the complications of the realistic
crystal structures and orbital degeneracy of VO$_2$ for a model
Hamiltonian, the Dimer Hubbard Model (DHM), that captures the key
competition between Mott localization due to Coulomb repulsion and
singlet dimerization, i.e. Peierls localization. This permits a
detailed systematic study that may clearly expose the physical
mechanisms at play.  Importantly, in our study the underlying lattice
{\em stays put}.  Therefore, we can directly address the issue whether
a purely electronic transition, having a bearing on the physics of
VO$_2$, exists in this model. The specific questions that we shall
address are the following: ({\it i}) Does this purely electronic model
predict a first order metal-insulator transition as a function of the
temperature {\em within the relevant parameter region}? ({\it ii})
What is the physical nature of the different states? ({\it iii}) Can
they be related to key available experiments?  These issues are
relevant, since if this basic model fails to predict an electronic MIT
consistent with the one observed in VO$_2$, then it would be mandatory
to include the lattice degrees of freedom.  In the present study we
shall provide explicit answers to these questions. We show that the
solution of the DHM brings the equivalent physical insight for VO$_2$
as the single band Hubbard model for V$_2$O$_3$, which is one of the
significant achievements of DMFT \cite{Georges1996,Kotliar2004}.

The dimer Hubbard model is defined as
\begin{align}
\label{model}
\nonumber
H=&[ -t \sum_{\langle i, j\rangle \alpha\sigma}
c^\dagger_{i\alpha\sigma} c_{j\alpha \sigma} +\ t_{\perp}
\sum_{i\sigma} c^\dagger_{i1\sigma} c_{i2\sigma} + h.c. ] \\
&+ \sum_{i\alpha} U n_{i\alpha\uparrow} n_{i\alpha\downarrow}
\end{align}
where $\langle i, j\rangle$ denotes n.n. lattice sites, $\alpha=
\{1,2\}$ denote the dimer orbitals, $\sigma$ is the spin, $t$ is the
lattice hopping, $t_{\perp}$ is the intra-dimer hopping, and $U$ is
the Coulomb repulsion.  For simplicity, we adopt a
semicircular density of states $\rho(\varepsilon)=
\sqrt{4t^2-\varepsilon^2}/(2\pi t^2) $.  The energy unit is set by
$t$=1/2, which gives a full bandwidth of 4$t$=2$D$=2, where $D$ is the
half-bandwidth.  This interesting model has surprisingly received
little attention, and only partial solutions have been obtained within
DMFT \cite{Moeller1998, Fuhrmann2006, Hafermann2009, Kancharla2007}.
The main results were the identification of the region of coexistent
solutions at moderate $U$ and small $t_\perp$ at $T$=0 using the iterated perturbation
theory (IPT)
approximation \cite{Moeller1998} and at finite $T$=0.025 by quantum
Monte Carlo\cite{Hirsch1983, Hirsch1986} (QMC)
\cite{Fuhrmann2006}. Here we obtain the detailed solution of the
problem paying special attention to the MIT and the nature of the
coexistent solutions.  We solve for the DMFT equations with
hybridization-expansion continuous-time quantum Monte Carlo (CT-QMC)
\cite{ctqmc_solver_werner2, Seth2015} and exact
diagonalization\cite{Georges1996}, which provide (numerically) exact
solutions.  We also adopt the IPT approximation \cite{Moeller1998},
which, remarkably, we find is (numerically) exact in the atomic limit
$t=0$, therefore provides reliable solutions of comparable quality as
in the single-band Hubbard model \cite{Georges1996}. Furthermore, IPT
is extremely fast and efficient to explore the large parameter space
of the model and provides accurate solutions on the real frequency
axis.  Extensive comparison between IPT and the CT-QMC is shown in the
Supplemental Material.  The DMFT equations provide for the exact
solution of the DHM in the limit of large lattice coordination and
have been derived elsewhere \cite{Moeller1998}. Here we quote the key
self-consistency condition of the associated quantum dimer-impurity
model,
\begin{equation}
{{\mathbf {G}}}^{-1}(i\omega_n)+ \bm{\Sigma}(i\omega_n) = \left( \begin{array}{cc}
i\omega_n  & -t_\perp \\
-t_\perp  &  i\omega_n \end{array} \right)
- t^2 {\bf G}(i\omega_n),
\label{eq:self}
\end{equation}
where $G_{\alpha,\beta}$ and $\Sigma_{\alpha,\beta}$ (with $\alpha,
\beta= 1,2$) are respectively the dimer-impurity Green's function and
self-energy. At the self consistent point these two quantities become
the respective local quantities of the {\em lattice}
\cite{Georges1996}. An important point to emphasize is that this
quantum dimer-impurity problem is {\it analogous} to that in the above
mentioned LDA+DMFT studies \cite{Biermann2005, Brito2015, Silke2008,
Silke1}. Therefore, strictly speaking, our methodology is a
Cluster-DMFT (CDMFT) calculation (cf Supplemental Material).

\begin{figure}
 \includegraphics[width=0.45\textwidth]{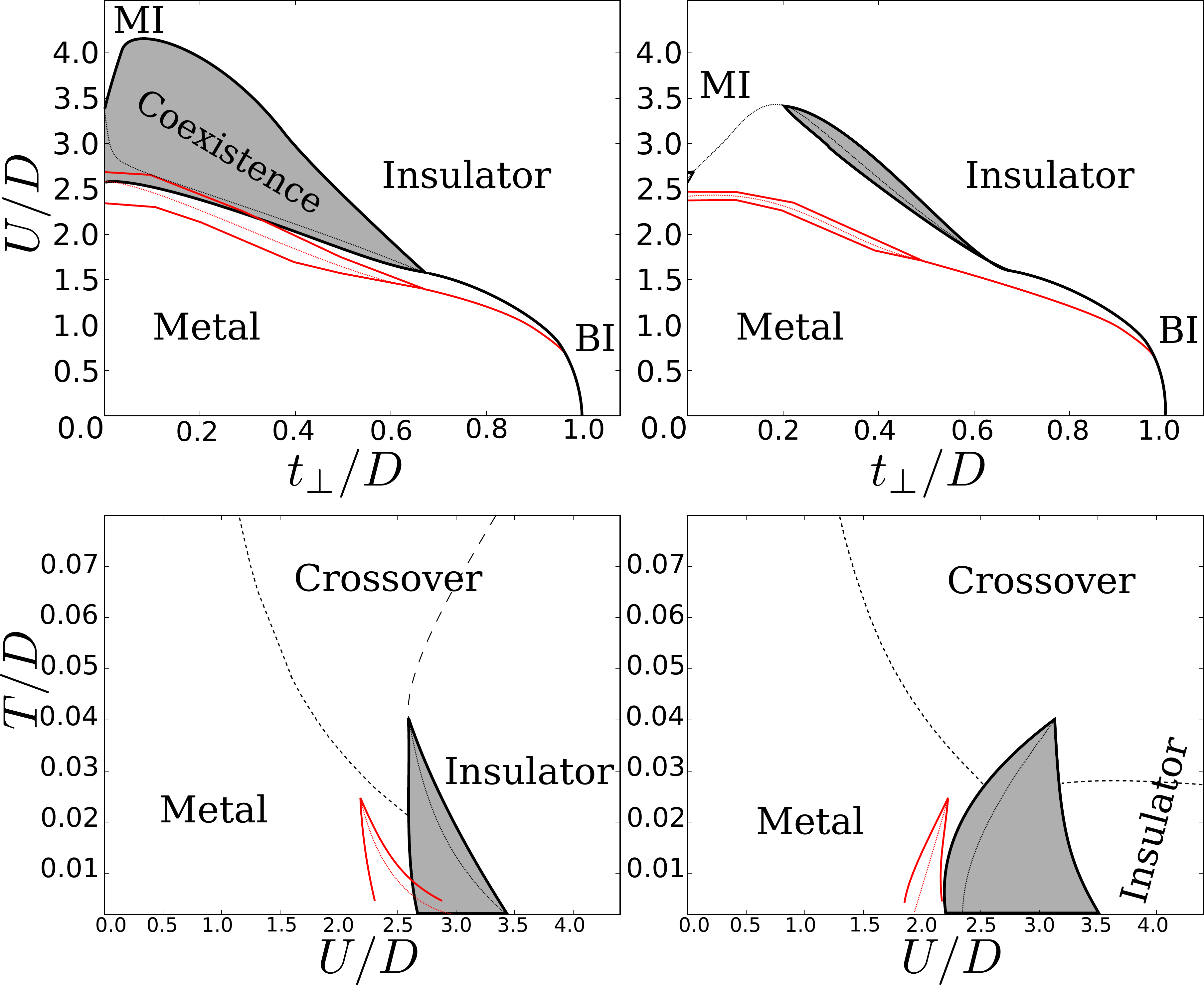}
\caption{Phase diagram showing the coexistence (grayed) of metal and
insulator states (black lines from IPT and red from QMC), where the
approximate position of the 1st order lines is indicated.  MI denotes
Mott insulator and BI bond insulator, the crossover regions have bad
metal behavior (see text and Ref.\onlinecite{Jacksa2015}).  Top panels
show $t_\perp$-$U$ plane. Left one shows lower temperatures $T$=0.001
(IPT) and 1/200 (CT-QMC), and right one shows higher temperature
$T$=0.03 (IPT) and 1/64 (CT-QMC).  Lower panels show the $U$-$T$
plane. Left one is for fixed $t_\perp$=0 ( i.e. single-band Hubbard
model), and right one for $t_\perp$=0.3 .}
\label{fig-phasediag}
\end{figure}

We start by establishing the detailed phase diagram, which we show in
Fig.~\ref{fig-phasediag}.  We observe that at low $T$ there is a large
coexistent region at moderate $U$ and $t_\perp$ below 0.6
\cite{Moeller1998}. This region gradually shrinks as $T$ is increased,
and fully disappears at $T\approx 0.04$. The lower panel shows the
phase diagram in the $U$-$T$ plane at fixed $t_\perp$.  At $t_\perp$=0
we recover the well known single-band Hubbard model result, where the
coexistent region extends in a triangular region defined by the lines
$U_{c1}(T)$ and $U_{c2}(T)$ \cite{Georges1996}. The triangle is tilted
to the left, which indicates that upon warming the correlated metal
undergoes a first order transition to a finite-$T$ Mott insulator.
This behavior was immediately associated to the famous 1st order MIT
observed in {\it Cr-doped} V$_2$O$_3$ \cite{Rozenberg1994,
Georges1996}, which has been long considered a prime example of a Mott
Hubbard transition \cite{Imada1998}. It is noteworthy that this
physical feature has remained relevant even in recent LDA+DMFT
studies, where the full complexity of the lattice and orbital
degeneracy is considered \cite{Hansmann2013, Lechermann2012}. This
underlines the utility of sorting the detailed behavior of basic model
Hamiltonians.  Significantly, as $t_\perp$ is increased in the DHM,
the tilt of the triangular region evolves towards the right.  This
signals that $t_\perp$ fundamentally changes the stability of the
groundstate. In fact, as shown in the lower right panel of
Fig.~\ref{fig-phasediag}, at $t_\perp$=0.3 we find that the MIT is
{\it reversed} with respect to the previous case. Namely, upon
warming, an insulator undergoes a 1st order transition to a (bad)
correlated metal at finite-$T$.  We may connect several features of
this MIT to VO$_2$, both, qualitative and semi-quantitatively.  We first
consider the energy scales and compare the parameters
of the DHM to those of electronic structure calculations. The LDA estimate of the bandwidth of the
metallic state of VO$_2$ is $\sim$2eV \cite{Biermann2005}, which
corresponds in our model to $4t$, hence $t$=0.5eV.  This is handy,
since from our choice of $t$=0.5, we may simply read the numerical
energy values of the figures directly in physical units (eV) and
compare to experimental data of VO$_2$.  We notice that the
coexistence region (with a 1st order transition line) extends up to
$T\approx$ 0.04 (eV) $\approx$ 400K, consistent with the experimental
value $\approx$ 340K.  We then set the value of $t_\perp$=0.3eV, that
approximately corresponds with LDA estimates for the (average)
intra-dimer hopping amplitudes (cf Supplemental Material)
\cite{Biermann2005, Lazarovits2010, Belozerov2012}. Thus, the
coexistence region is centered around $U\approx 2.5-3 eV$, consistent
with the values adopted in the LDA+DMFT studies \cite{Biermann2005,
Lazarovits2010}.

\begin{figure}
 \includegraphics[width=0.5\textwidth]{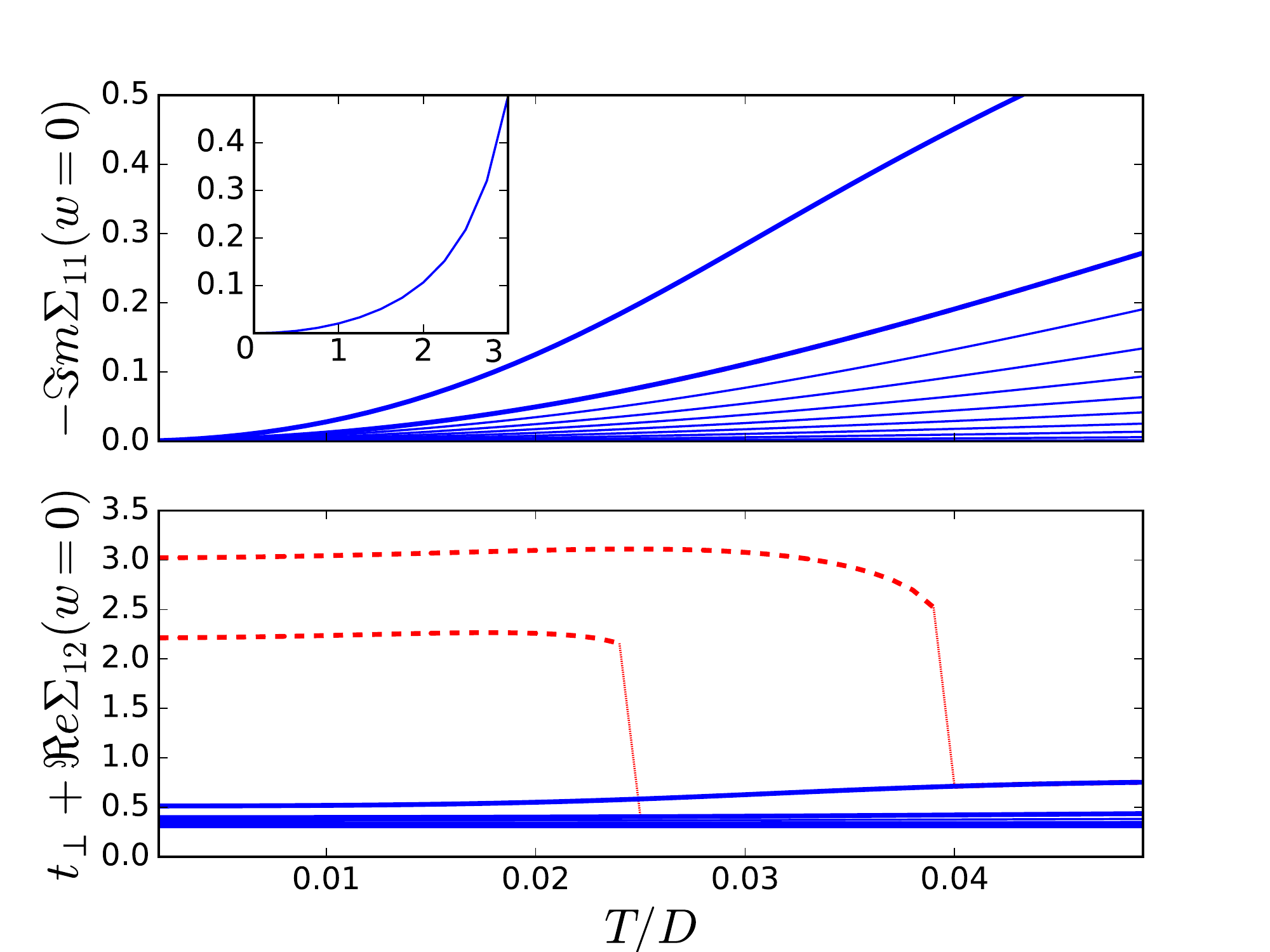}
\caption{Top: The scattering rate Im[$\Sigma_{11} (\omega=0)]$ for the
metal (solid) at fixed $t_\perp$=0.3 values of $U$ from 0 to 3 in
steps of 0.5 (upwards). The experimentally relevant values $U$=2.5 and
3, are highlighted with thick lines. Inset shows the $U$ dependence at
fixed $T$=0.04.  Bottom: The effective intra-dimer hopping
${t_\perp}^{eff}$= $t_\perp$+Re$[\Sigma_{12}](0)$ (bottom) as a
function of $T$ for the same parameters as the top panel.  Metal
states are in solid (blue) lines and the insulator in dashed (red)
lines for $U$=2.5 and 3. The calculation are done with IPT.}
\label{fig-sigma}
\end{figure}

We can make further interesting connections with experiments in
VO$_2$. The metallic state is unusual and it can be characterized as a
{\it bad metal}.  Namely, a metal with an anomalously high scattering
rate that approaches (or may violate) the Ioffe-Regel limit
\cite{badmetals}. In Fig.~\ref{fig-sigma} we show the imaginary part
of the diagonal self-energy, whose y-axis intercept indicates the
scattering rate (i.e. inverse scattering time). At $T\approx$ 0.04
(i.e.$\sim$400K) we observe a large value of the intercept, of order
$\sim t$=1/2, which signals that the carriers are short lived
quasiparticles.  In fact VO$_2$ has such an anomalous metallic state
\cite{Qazilbash2007}. This anomalous scattering is likely the origin
of the surprising observation that despite the fact that the lattice
structure has 1D vanadium chains running along the c-axis, the
resistivity is almost isotropic, within a mere factor of 2
\cite{Bongers1965}.  It is noteworthy that this lack of anisotropy
observed in electronic transport experiments provides further
justification for our simplified model of a lattice of dimers.  This
bad metal behavior is a hallmark of Mottness \cite{mottness,
Jacksa2015} and also indicates that the MIT in VO$_2$ should be
characterized as a Mott transition\footnote{See Supplementary Material
section 5}.  Additional insights on the mechanism driving the
transition can be obtained from the behavior of the off-diagonal
(intra-dimer) self-energy $\Sigma_{12}(\omega_n)$. From
Eq.~(\ref{eq:self}), we observe that the intra-dimer hopping amplitude
is effectively renormalized as ${t_\perp}^{eff} = t_\perp + \Re
e[\Sigma_{12}](0)$. In Fig.~\ref{fig-sigma} we show the behavior of
this quantity across the transition. We see that in the metallic state
it remains small, while it becomes large ($\gg t_\perp$) at low
$T$ in the insulator \cite{Brito2015, Silke2008, Silke1}.  The
physical interpretation is transparent. In the correlated metal, the
two dimer sites are primarily Kondo screened by their lattice
neighbors, as in the single band Hubbard model each one forms a heavy
quasiparticle band. Then these two bands get split into a bonding and
anti-bonding pair by the small $t_\perp$. Hence, the low energy
electronic structure is qualitatively similar to the non-interacting
one, with a larger effective mass.  As $T$ is lowered, the dramatic
increase in $\Re e[\Sigma_{12}](0)$ when the Mott gap opens at the 1st
order transition signals that the intra-dimer interaction is boosted
by ${t_\perp^{eff}} \sim \Re e[\Sigma_{12}]$. Unlike the one-band
Hubbard model, here the finite $t_\perp$ permits a large energy gain
in the Mott insulator by quenching the degenerate entropy. This
mechanism, already observed in other cluster-DMFT models
\cite{parcollet_cluster_2004, Park2008, Balzer2009}, stabilizes the
insulator within the coexistence region, leading to the change in the
tilt seen in Fig.\ref{fig-phasediag}.  Another way of rationalizing
the transition is that at a critical $U-$dependent $t_\perp$ the Kondo
screening in the metal breaks down in favor of the local dimer-singlet
formation in the insulator.  In this view, the large gap opening may
be interpreted as a $U-$driven enhanced band splitting $\propto
{2t_\perp^{eff}(U)}$.

\begin{figure}
 \includegraphics[width=0.45\textwidth]{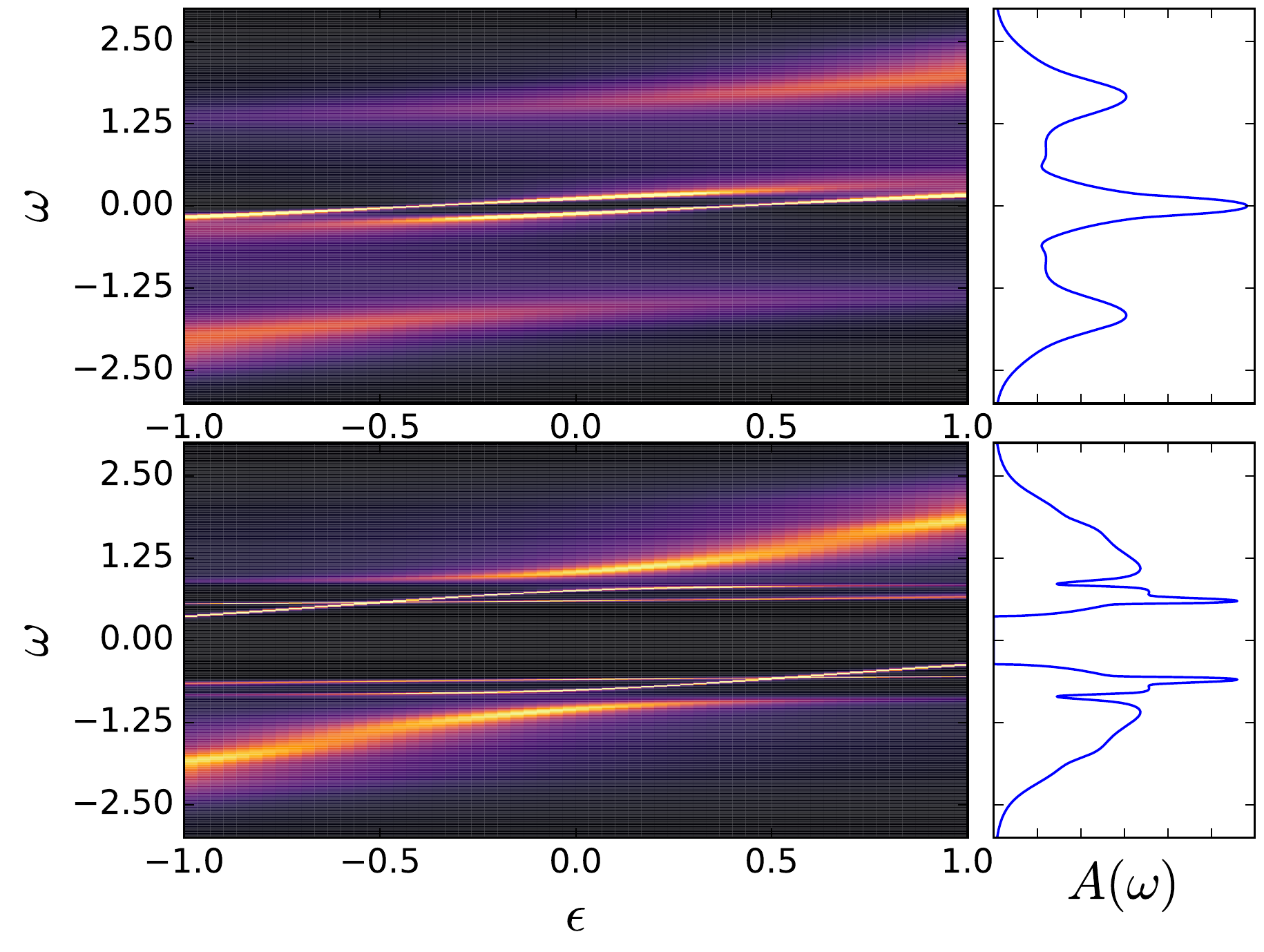}
\caption{Electronic dispersion for the metal (left top) and insulator
(left bottom) in the coexistence region for parameter values
$t_\perp$=0.3, $U$=2.5 and $T$=0.01.  Right panels show the respective
DOS($\omega$). The calculations are done with IPT(cf. Sup. Mat.)}
\label{fig-arpes}
\end{figure}

Further detail is obtained from the comparison of the electronic
structure of the metal and the insulator within the coexistence region
\footnote{The electronic structure results from IPT, CT-QMC and exact
diagonalization show all good agreement, see Supplemental Material}.
In the correlated metallic state shown in Fig.~\ref{fig-arpes} we find
at high energies ($\sim \pm U/2$) the incoherent Hubbard bands, which
are signatures of Mott physics.  At lower energies, we also observe a
pair of heavy quasiparticle bands crossing the Fermi energy at
$\omega$=0. Consistent with our previous discussion, this pair of
quasiparticle bands can be though of as the renormalization of the
non-interacting bandstructure.  Significantly, as we shall discuss
later on, this feature may explain the puzzling optical data of
Qazilbash et al. \cite{Qazilbash2007} within the MIT region of VO$_2$,
which has remained unaccounted for so far.  Unlike the single-band
Hubbard model, the effective mass of these metallic bands does not
diverge at the MIT at the critical $U$, even at $T$=0. In fact, the
finite $t_\perp$ cuts off the effective mass divergence as expected in
a model that incorporates spin-fluctuations.  In fact the DHM may be
considered \cite{Moeller1998} the simplest non-trivial cluster DMFT
model.  It is interesting to note that the realistic values $U=2.5$
and $t_\perp$ =0.3 lead to Hubbard bands at $\approx \pm 1.5$eV and a
quasiparticle residue $Z\approx 0.4$, both consistent with
photoemission experiments of Koethe et al. \cite{Koethe2006}.

In Fig.~\ref{fig-arpes} we also show the results for insulator
electronic dispersion at the same values of the parameters. The
comparison of the insulator and the metal illustrate the significant
changes that undergo at the 1st order MIT.  We see that the metallic
pair of quasiparticle bands suddenly open a large gap.  More
precisely, in contrast to the one-band case, here the Hubbard bands
acquire a non-trivial structure, with sharp bands coexisting with
incoherent ones.
The coherent part dispersion can be traced to those of a lattice of
singlet-dimers (see Sup. Mat.).  Hence, the insulator can be
characterized as a novel type of Mott-singlet state where the Hubbard
bands have a mix character with both coherent and incoherent
electronic-structure contributions.
It is also interesting to note that the gap in the density of states
is $\Delta \approx$0.6eV, again consistent with the photoemission
experiments \cite{Koethe2006}.

\begin{figure}
 \includegraphics[width=0.45\textwidth]{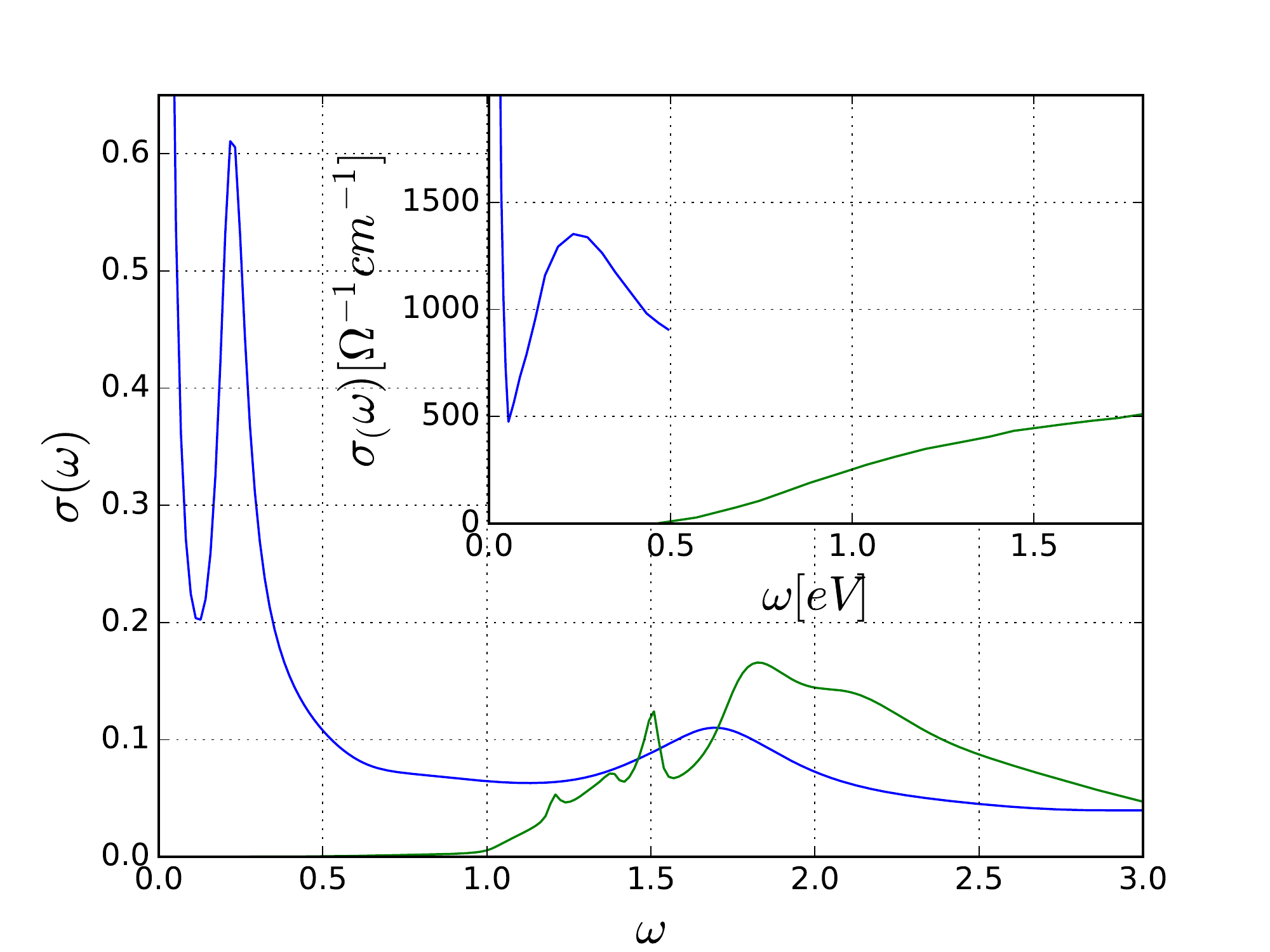}
\caption{The optical conductivity $\sigma(\omega)$ of metal and
insulator within the coexistence region for parameters $t_\perp$=0.3,
$U$=2.5 and $T$=0.01.  The calculations are done with IPT.  Inset: The
experimental optical conductivity adapted from
Ref.~\onlinecite{Qazilbash2007}.}
\label{fig-AC}
\end{figure}

In order to gain further insight and make further contact with key
experiments, we now consider the optical conductivity response
$\sigma(\omega)$ within the MIT coexistence region.  A set of
remarkable data was obtained in this regime by Qazilbash et
al. \cite{Qazilbash2007}, bearing directly on the issue of the driving
force behind the transition. They systematically investigated the
$\sigma(\omega)$ as a function of $T$ using nano-imaging spectroscopy.
They clearly identified within the $T$ range of the MIT the electronic
coexistence of insulator and metallic regions, characteristic of a 1st
order transition. A crucial observation was that upon warming the
insulator in the M1 phase, metallic puddles emerge with a
$\sigma(\omega)$ that was significantly different from the signal of
the normal metallic R phase.  Thus, the data provided a strong
indication of a purely electronic driven transition.  Regarding this
point we would like to mention also the works of Arcangeletti et
al. \cite{Arcangeletti2007} and Laverock et al. \cite{Laverock2014}
that reported the observation of metallic states within the monoclinic
phase under pressure and strain, respectively. Coming back to the
experiment of Qazilbash et al., a key point that we want to emphasize
here is that $\sigma(\omega)$ in the putative M1-metallic state was
characterized by a intriguing mid-infra-red (MIR) peak $\omega_{mir}
\approx$ 1800 cm$^{-1}=0.22$ eV, whose origin was not understood.
From our results on the electronic structure within the coexistence
region, we find a natural interpretation for the puzzling MIR peak: It
corresponds to excitations between the split metallic quasiparticle
bands. Since they are parallel, they would produce a significant
contribution to $\sigma(\omega)$, which enabled its detection. In
Fig.~\ref{fig-AC} we show the calculated optical conductivity response
(see Sup. Mat. section 7) that corresponds to the spectra of
Fig.~\ref{fig-arpes}.  In the metal we see that, in fact, a prominent
MIR peak is present at $\omega_{mir} \approx$ 0.22 eV, in excellent
agreement with the experimental value. On the other hand, the optical
conductivity of the insulator shows a maximum at $\omega_{ins} \approx
2$ eV in both, theory and experiment. Moreover, we also note the good
agreement of the relative spectral strengths of the main features in
the two phases.

In conclusion, we showed that the detailed solution of the dimer model
treated within DMFT can account for a number of experimental features
observed in VO$_2$. The minimal model has an impurity problem which is
analogue to that of LDA+DMFT methods, yet the simplicity of this
approach allowed for a detailed solution that permitted a transparent
understanding of many physical aspects of the electronic first order
transition in this problem.  It exposes a dimer-Mott-transition
mechanism, where the effective intra-dimer exchange is controlled by
correlations, it is weakened in the metal and strongly enhanced in the
Mott insulator. In the metal, this leads to a pair of split
quasiparticle bands, which then in the insulator further separate, to
join and coexist with the usual incoherent Hubbard bands. Despite the
simplicity of our model, we made semi-quantitative connections to
several experimental data in VO$_2$, including a crucial optical
conductivity study within the 1st order transition, that remained
unaccounted for.  Our work, sheds light on the long-standing question
of the driving force behind the metal-insulator transition of VO$_2$
highlighting the relevance of the Mott mechanism. The present approach
may be considered the counterpart for VO$_2$, of the DMFT studies of
the Mott transition in paramagnetic Cr-doped V$_2$O$_3$.

%\section{Acknowledgments}

{\it Acknowledgments} We thank I. Paul, S. Biermann, G. Kotliar and
H-T Kim for helpful discussions.  This work was partially supported by
public grants from the French National Research Agency (ANR), project
LACUNES No ANR-13-BS04-0006-01, the NSF DMR-1005751 and DMR-1410132.

\putbib[biblio,vo2-arxivNotes]

\end{bibunit}

%%%%%%%%%% Merge with supplemental materials %%%%%%%%%%
%\pagebreak
\widetext
\begin{center}
\textbf{\large Supplementary Material}
\end{center}
%%%%%%%%%% Merge with supplemental materials %%%%%%%%%%
%%%%%%%%%% Prefix a "S" to all equations, figures, tables and reset the counter %%%%%%%%%%
%\setcounter{equation}{0}
%\setcounter{figure}{0}
%\setcounter{table}{0}
%\setcounter{page}{1}
\makeatletter
\renewcommand{\theequation}{S\arabic{equation}}
\renewcommand{\thefigure}{S\arabic{figure}}
\renewcommand{\bibnumfmt}[1]{[S#1]}
\renewcommand{\citenumfont}[1]{S#1}
%%%%%%%%%% Prefix a "S" to all equations, figures, tables and reset the counter %%%%%%%%%%
\begin{bibunit}[unsrtnat]

\section{Dimer Hubbard Model (DHM) representation}
\label{sec:orgheadline1}
\begin{figure}[htb]
\centering
\includegraphics[width=7cm]{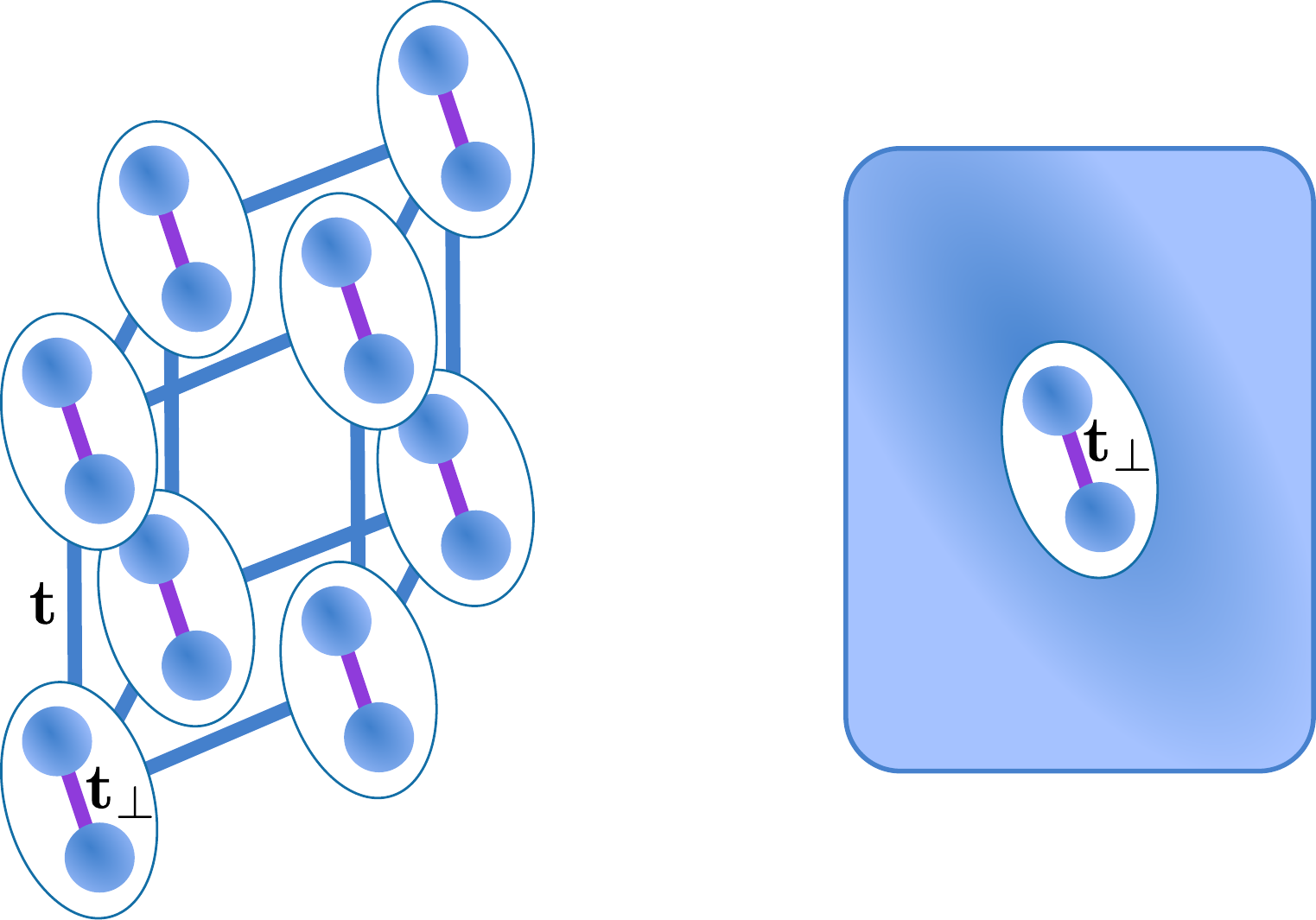}
\caption{Left panel: schematic representation of the lattice Hamiltonian. Right panel: Asociated self-consistent quantum impurity problem, where the dimer is embedded in a self-consistent medium. \label{fig:dimer_lattice_representation}.}
\end{figure}

In figure \ref{fig:dimer_lattice_representation} we show on the left
panel the schematic representation of our model Hamiltonian. The blue
lines correspond the inter-dimer hopping matrix \(\mathbf{t}\) and the purple
line the intra-dimer hopping \(\mathbf{t_\perp}\). For simplicity the model is
depicted in 3D, but it is mathematically formulated in the limit of a
large coordination lattice. The right hand side panel shows the
corresponding DMFT quantum impurity problem, where the dimer is
embedded in a self consistent medium.

\section{Validation of IPT against the exactly solvable isolated dimer limit (ie atomic limit of the lattice model)}
\label{sec:orgheadline2}
\begin{figure}[h]
\centering
\subfloat{\includegraphics[width=0.3\textwidth]{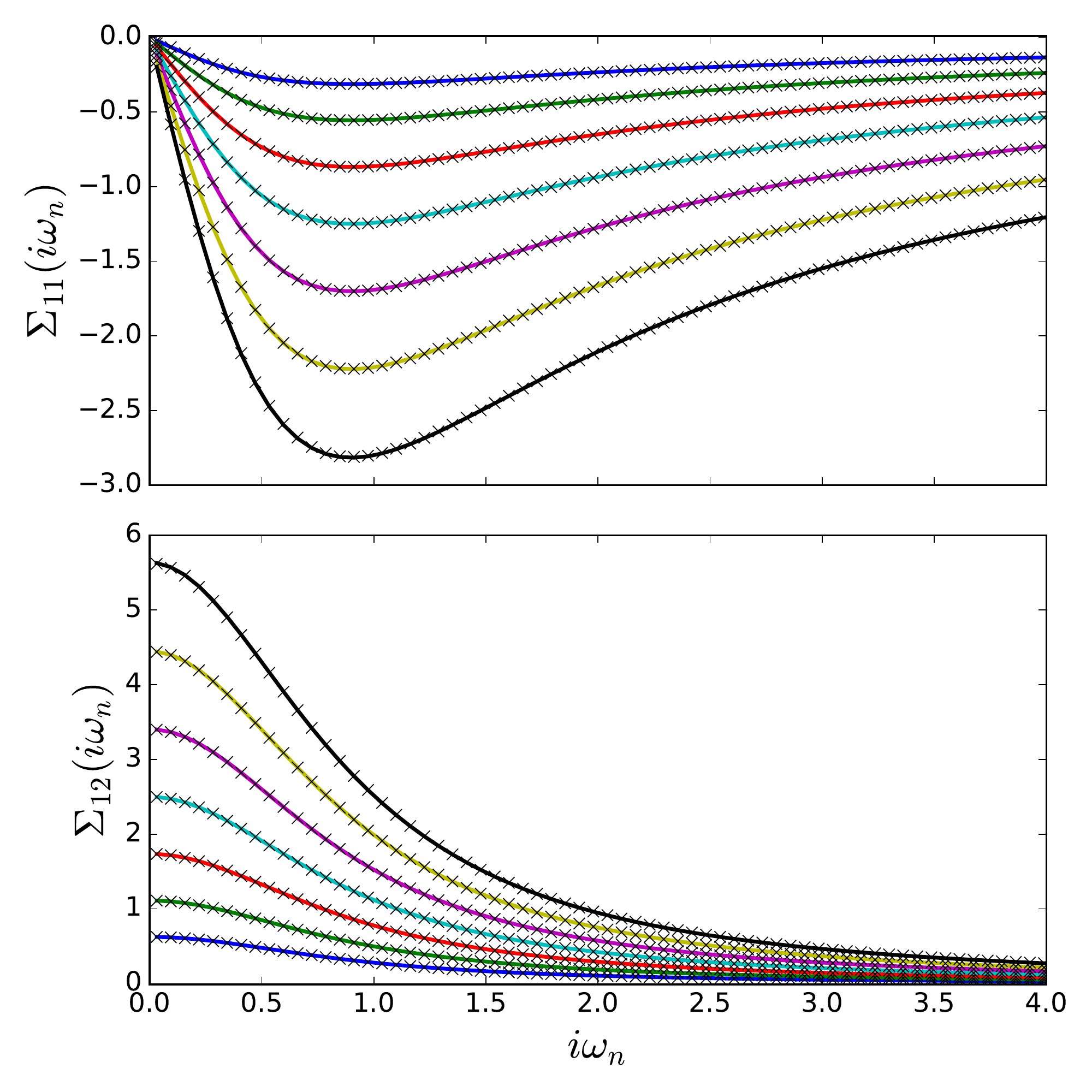}}
\subfloat{\includegraphics[width=0.3\textwidth]{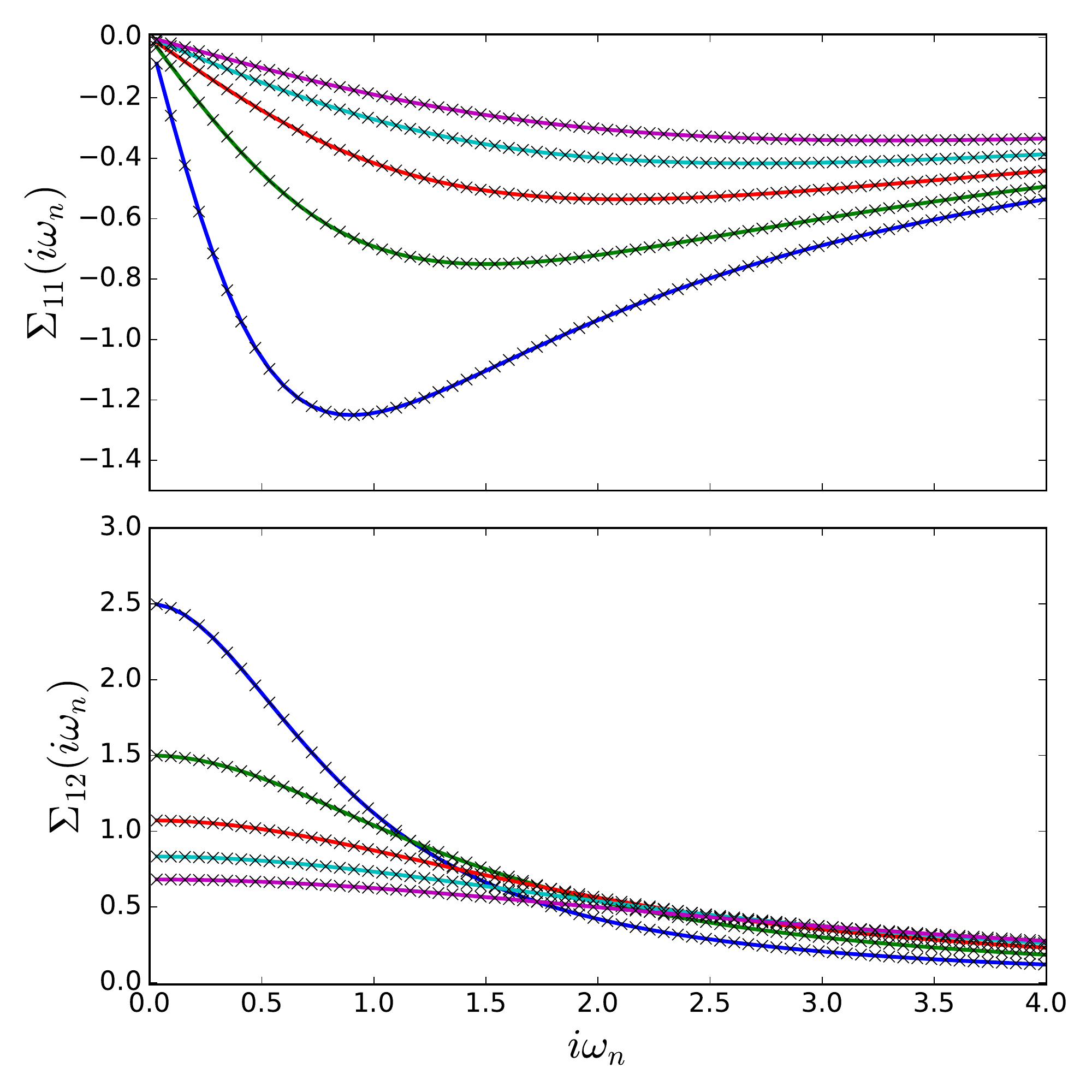}}
\subfloat{\includegraphics[width=0.3\textwidth]{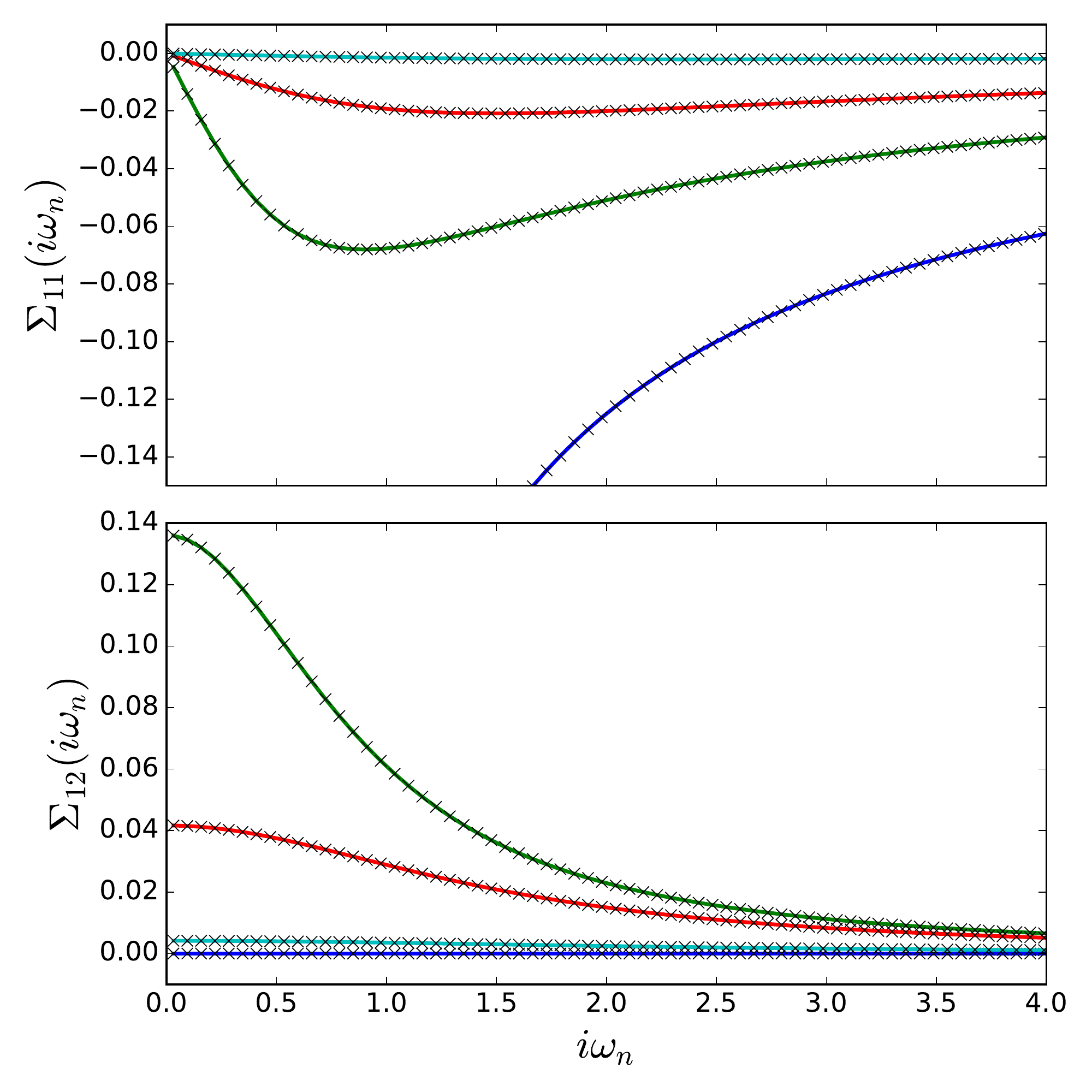}}
\caption{Comparison between the exact Self-Energy of an isolated dimer(solid lines) and the approximation from IPT(black crosses) (Left panel)Self-Energy for various values of $U$ and fixed $t_\perp=0.3$. $U=1.5$ blue, $U=2$ green, $U=2.5$ red, $U=3$ cyan, $U=3.5$ purple, $U=4$ yellow, $U=4.5$ black. (Center) Various values of $t_\perp$ and fixed $U=3$. $t_\perp=0.3$ blue, $t_\perp=0.5$ green, $t_\perp=0.7$ red, $t_\perp=0.9$ cyan, $t_\perp=1.1$ purple. (Right) Various values of $U$ and $t_\perp$. ($U=1$, $t_\perp=0$) blue, ($U=0.7$, $t_\perp=0.3$) green, ($U=0.5$, $t_\perp=0.5$) red, ($U=0.2$, $t_\perp=0.8$) cyan.\label{fig:IPT_dimer_exact}}
\end{figure}

We numerically demonstrate that the IPT method exactly captures the
atomic limit of the lattice model. The fact that a perturbative
calculation captures the atomic limit (ie, \(U/t \to \infty\)) is
not to be expected, but, interestingly enough, is analogous to the
well known property of the IPT solution of the one band Hubbard
model. This is shown in figure \ref{fig:IPT_dimer_exact} where we
compare the exact Self-Energy of the atomic limit (isolated dimer)
with the respective IPT solution.

\section{Comparison of IPT with numerical solutions (CT-QMC and Exact Diagonalization)}
\label{sec:orgheadline5}
\subsection{Solutions in the Matsubara axis}
\label{sec:orgheadline3}
\begin{figure}\centering
\subfloat{\includegraphics[width=0.4\textwidth]{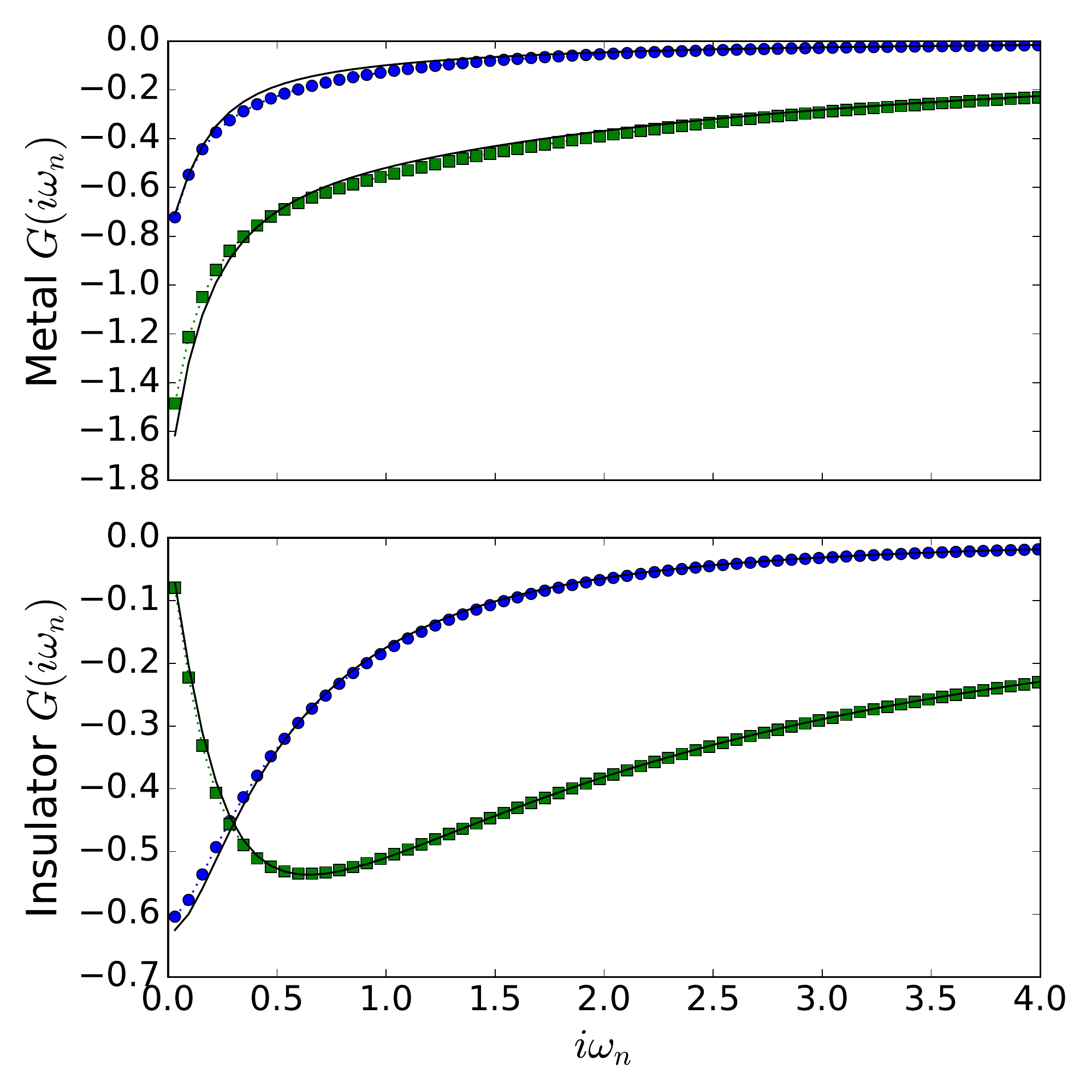}}
\subfloat{\includegraphics[width=0.4\textwidth]{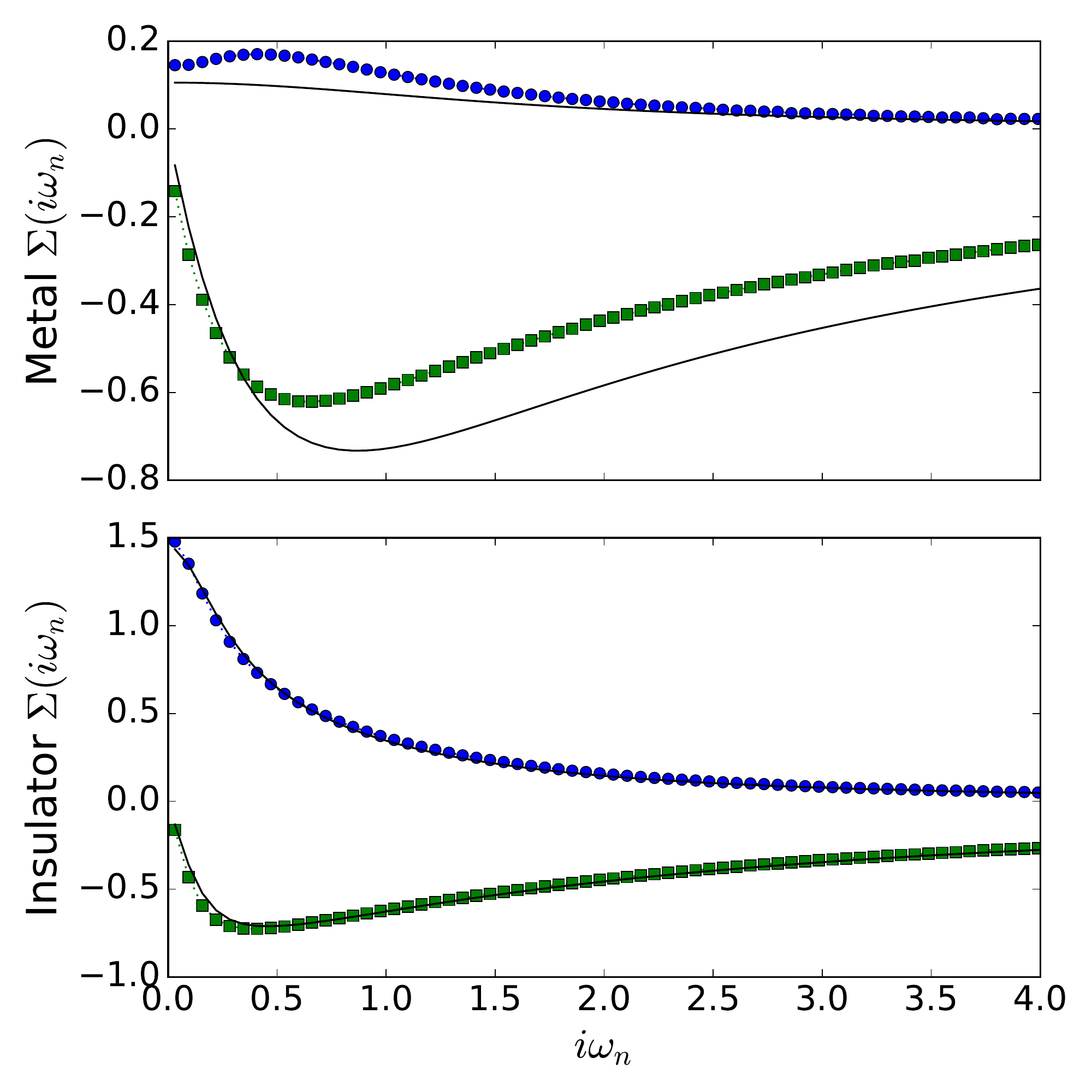}}
\caption{Comparison of the (numerically) exact CT-QMC solution and IPT within
the coexistence region for metal (top) and insulator (bottom).
Left panels: the local Im[$G_{11}$] (green) and Re[$G_{12}$] (blue).
Right panels: the local Im[$\Sigma_{11}$] (green) and Re[$\Sigma_{12}$] (blue).
Circular and square symbols are CT-QMC data at $U$=2.15 and $t_\perp$=0.3,
and black solid lines are IPT at $U$=2.55 (metal) and $U$=2.21 (insulator) with $t_\perp$=0.3.
Notice that, as in the one band Hubbard model case, the best quantitative agreement
between QMC and IPT is found for values of $U$ that are slightly different.
\label{fig:CT-QMC-IPT-coex}}
\end{figure}

The most stringent test for the comparison is done within the
coexistence region, since there the structure of the Green's functions
and Self-Energies are very non-trivial. Figure
\ref{fig:CT-QMC-IPT-coex} shows the numerically exact CT-QMC Green's
functions and Self-Energies in the Matsubara axis. The data is shown
along with fits obtained from the IPT solution at suitably close
values of the parameters.

\begin{figure}\centering
\subfloat{\includegraphics[width=0.4\textwidth]{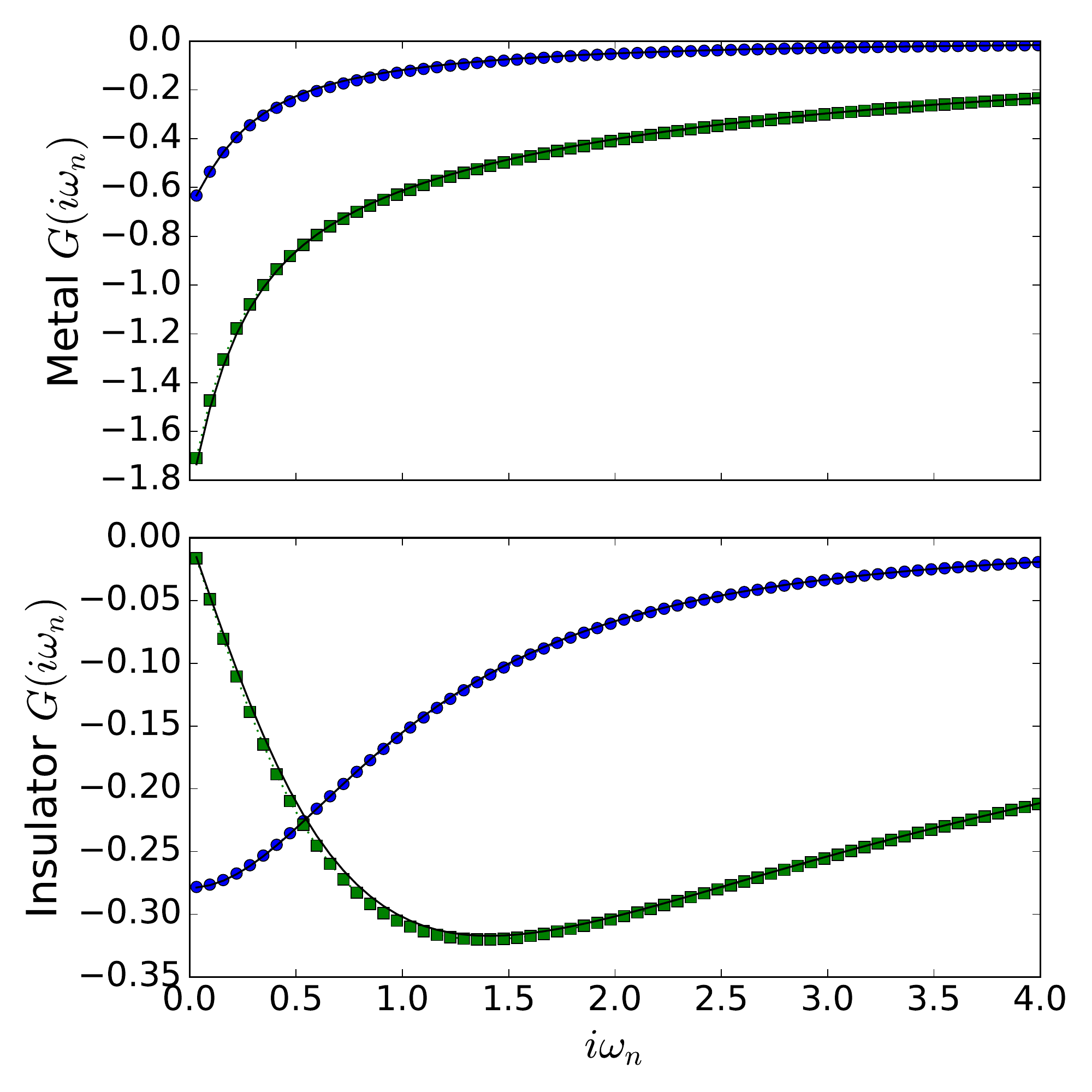}}
\subfloat{\includegraphics[width=0.4\textwidth]{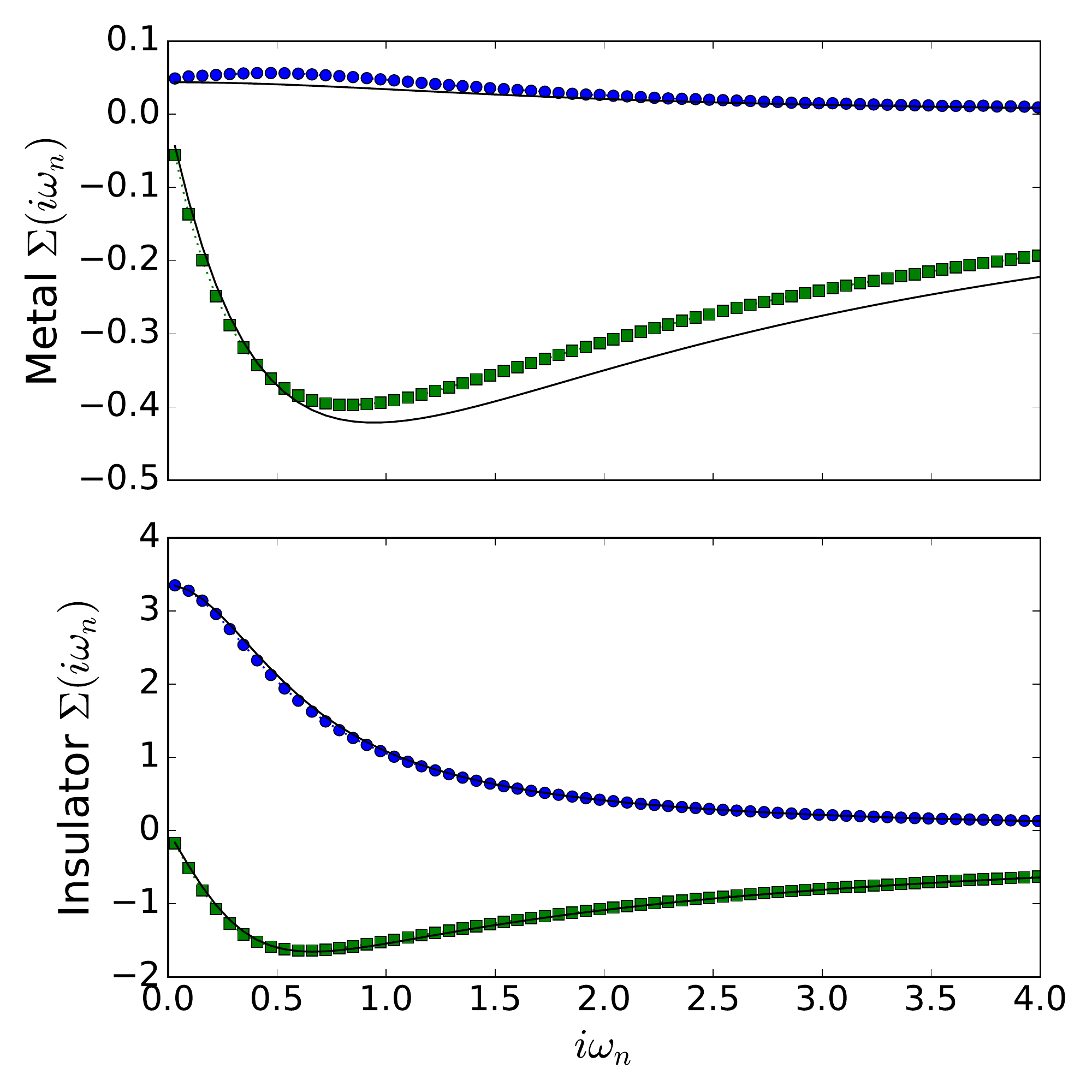}}
\caption{Comparison of the (numerically) exact CT-QMC solution and IPT away from
the coexistence region for metal (top) and insulator (bottom).
Left panels: the local Im[$G_{11}$] (green) and Re[$G_{12}$] (blue).
Right panels: the local Im[$\Sigma_{11}$] (green) and Re[$\Sigma_{12}$] (blue).
Circular and square symbols are CT-QMC data at $U$=1.8 (metal) and 3.3 (insulator),
and black solid lines are IPT at $U$=2 (metal) and 3.35 (insulator). The intra-dimer
hopping is always fixed at $t_\perp$=0.3.
Notice that, as in the one band Hubbard model case, the best quantitative agreement
between QMC and IPT is found for values of $U$ that are slightly different.
\label{fig:CTQMC-ipt-benign}}
\end{figure}

The very good agreement between QMC and IPT is also found in the whole
phase diagram. This is shown in figure \ref{fig:CTQMC-ipt-benign}
where IPT and CT-QMC are compared at two parameter values away from
the coexistence region.

\subsection{Electronic Structure}
\label{sec:orgheadline4}
Within DMFT, the bandstructure is obtained as a function of
the single particle energy \(\epsilon\), which in the semicircular DOS
lattice adopted here has a simple linear dispersion
\cite{Georges1996}. Hence at \(U=0\) the non-interacting bandstructure
are two parallel linear bands split by \(2t_\perp\) ( i.e. the bonding
and anti-bonding bands).

\begin{figure}\centering
\subfloat{\includegraphics[width=0.3\textwidth]{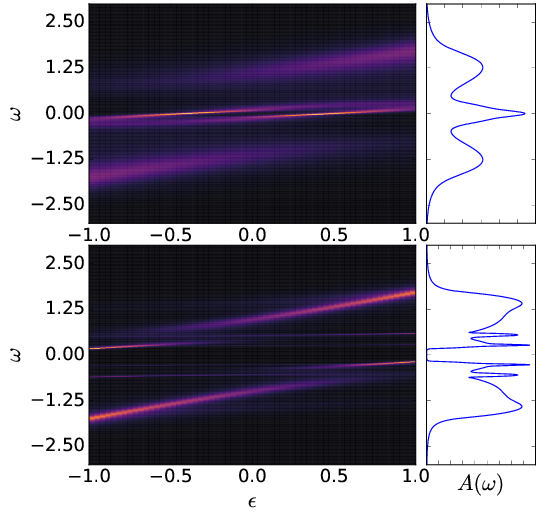}}
\subfloat{\includegraphics[width=0.3\textwidth]{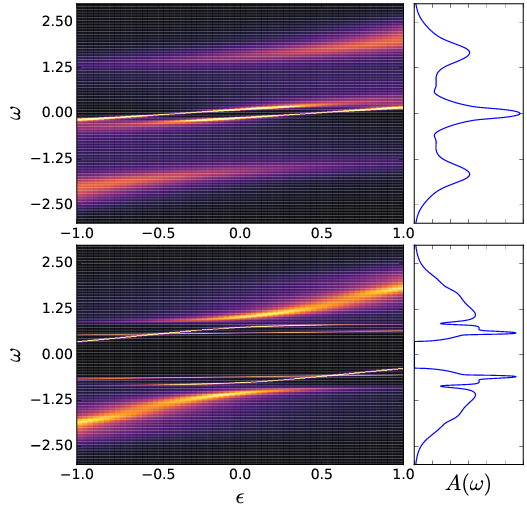}}
\subfloat{\includegraphics[width=0.3\textwidth]{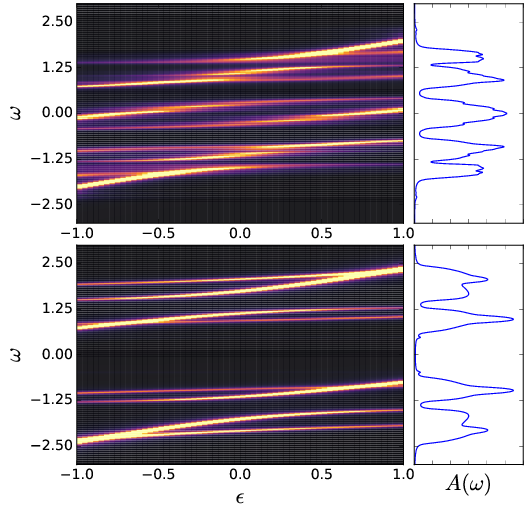}}
\caption{Comparison of the electronic structures in the Metal and Insulator phases at $t_\perp=0.3$. Left CT-QMC in the coexistence $U=2.15$, center IPT taken from main text $U=2.5$, right results from ED metal at $U=1.8$, insulator at $U=3$ \label{fig:electronic_structures}}
\end{figure}

In figure \ref{fig:electronic_structures} we show the electronic
dispersion (ie, \(\epsilon\) and \(\omega\) resolved density of states
\(A(\epsilon,\omega)\)) of the metallic and insulating states obtained
by the IPT, CT-QMC and ED methods, respectively. As seen in the
figure, all the three methods provide the key qualitative features
that are discussed in the text. Namely, the split quasiparticle bands in
the metal and in the insulating states.

\section{Choice of parameters in the Dimer lattice}
\label{sec:orgheadline6}
The dimer Hubbard Model has very few parameters inter-dimer (or
lattice) hopping \(t\), intra-dimer hopping \(t_\perp\), local Coulomb
repulsion \(U\), and the temperature. The lattice hopping provides the
bandwidth of our model Hamiltonian \(W=4t\)
\cite{Georges1996,Moeller1998}. In the case of VO\(_{\text{2}}\) reference LDA
calculations agree that both, \(e_g\) and \(a_{1g}\) bands have an
approximate bandwidth of 2eV
\cite{Biermann2005,Belozerov2012,Brito2015,Lazarovits2010}, hence we have
set \(t=0.5eV\).

The intra-dimmer hopping quoted in ref. \cite{Biermann2005} is
\(\sim\) 0.68eV, which is about a factor of 2 larger than our adopted
value 0.3eV. However, that value concerns solely the \(a_{1g}\)
(d\(_{\text{xy}}\)) orbital. There are in fact also two additional
intra-dimer hopping amplitudes (associated to e\(_{\text{g}}\)
states), which are 0.22eV and a smaller value. The values for the
hopping amplitudes found in later works
\cite{Lazarovits2010,Belozerov2012,Brito2015} are consistent with this
findings. Since we are considering a unit cell with two sites and one
orbital each, we have a single intra-dimer hopping
parameter. Therefore, and in the spirit of a mean field theory, we
chose a value that is the approximate average of the 3 hopping
amplitudes.

In regard to the value of \(U\), we adopted the value of \(U=2.5eV\) for
the semi-quantitative comparison with experiments. This is quite
consistent with the values considered in \cite{Lazarovits2010}, who
systematically explored the range of \(U=2.2\) to 3.5eV, and found the
MIT between 2.7 and 3eV at the lowest temperatures considered. The
values adopted in ref. \cite{Biermann2005,Belozerov2012} are \(U=4eV\) and
\(J=0.68\). These values are in fact higher than ours, however, Biermann
et al dedicate the last paragraph of their Letter to discuss their
choice of the value of \(U\) and \(J\). They mention that smaller values,
such as \(U=2eV\) are also compatible with their findings. This is also
consistent with our choice of \(U=2.5eV\).

\section{Mott insulator in the Dimer lattice}
\label{sec:orgheadline8}
\begin{figure}[ht]
  \centering
  \includegraphics[width=0.9\textwidth]{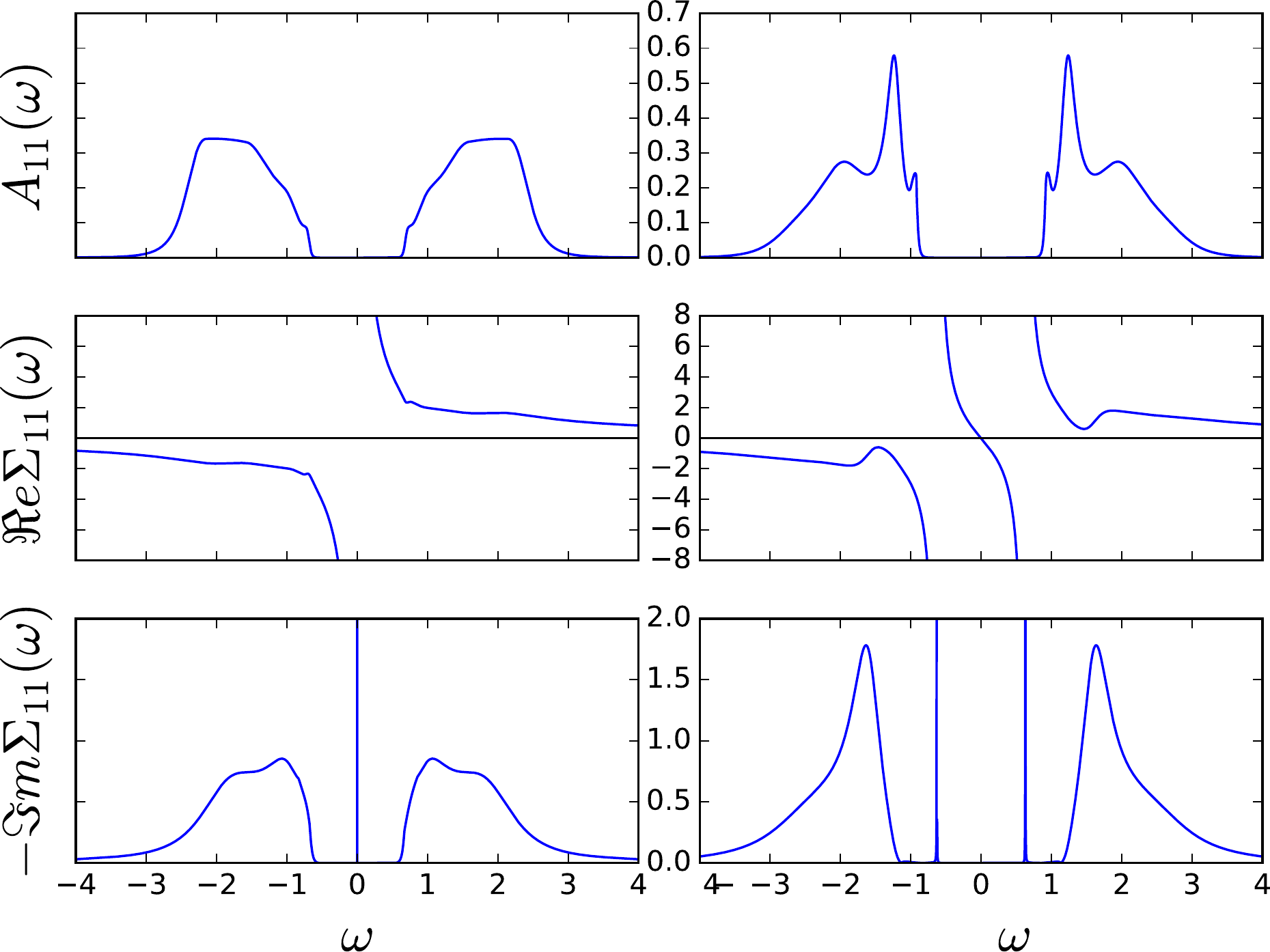}
  \caption{Local Spectral function(top Panel), Real Part of the
Self-Energy(middle) and minus Imaginary part of the self-energy. Left
column corresponds to the known DMFT solution of the single band
Hubbard model($t_\perp=0$). The right column corresponds to the dimer
lattice at $t_\perp=0.3$. For both figures
$U=3.5$\label{fig:IPT_dimer_Mott_gap} }
\end{figure}

The Mott insulating state is signaled by the divergence of the
Self-Energy in the Mott gap. This divergence is clearly visible on the
real axis in figure \ref{fig:IPT_dimer_Mott_gap}(middle and bottom
rows). In the left column we show the well known case of a single-band
Hubbard model (\(t_\perp=0\)) with particle-hole symmetry. In this case
the Self-Energy diverges at zero frequency, and this is therefore
visible in the Matsubara axis too. eg. in figure
\ref{fig:IPT_dimer_exact} above (the \(t_\perp=0\) blue squares on the
top most right panel). In the case of the dimer lattice
(\(t_\perp=0.3\)), with two atoms per unit cell, the dimerization splits
the divergence of the Self-Energy into two poles, symmetric around
zero frequency, as displayed on the right right column on figure
\ref{fig:IPT_dimer_Mott_gap}. In this case the divergence is no more visible
on the Matsubara axis, and in particular \(\Im m \Sigma_{11}\) goes
smoothly to zero as \(\omega \rightarrow 0\).

\subsection{The electronic structure of the dimer model}
\label{sec:orgheadline7}

\begin{figure}[htb]
\centering
\includegraphics[width=.9\linewidth]{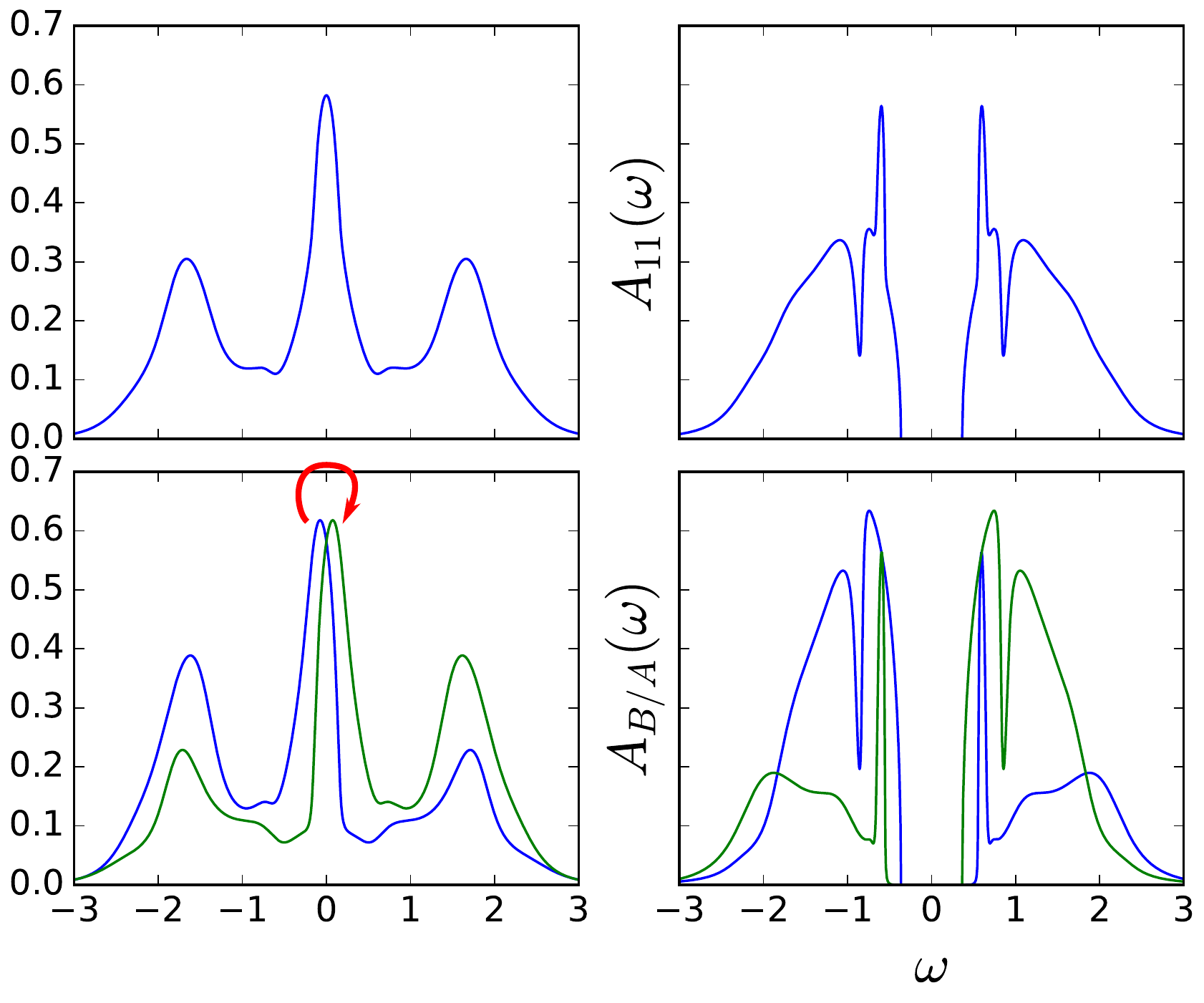}
\caption{Top panel shows the DOS in the atomic-site basis (from Fig.3 of manuscript) This DOS is doubly degenerate (the same at each of the two atoms of the unit cell). Bottom panel show the DOS in the bonding (blue) and antibonding (green) combinations, respectively. The average of the bottom two DOS gives the top one. Red arrow in bottom left panel indicates the origin of th MIR peak which appears at \(\omega \sim 0.22 eV\) (cf. fig. \ref{fig:decomposed_optcond}) \label{fig:IPT_ab_spectra_character}}
\end{figure}

The dimer Hubbard Model is not a single band model despite being a
single orbital per site system. Since it has two sites (the dimer) in
the unit cell, there are two bands. The two sites are related by
symmetry thus the two bands are degenerate in the atomic-site
representation but are distinct in the bonding/antibonding basis. The
physics of the system is independent of the choice of the
representation. To further clarify this point we show in
figure \ref{fig:IPT_ab_spectra_character} the DOS from our model
calculation from Fig. 3 of the manuscript in the atomic-site basis,
along with the bonding and antibonding DOS. The average of the latter
two gives the former one (which has degeneracy 2).

\section{Electronic structure of the atomic (isolated dimer) limit of the lattice model}
\label{sec:orgheadline9}

In order to identify the electronic structure of the insulator state
of the model, we obtained the corresponding quantity in the atomic
limit, which can be analytically solved in the real frequency axis
(the impurity model is an isolated dimer). In this limit the lattice
Green's function is obtained as,

\begin{equation}
\label{eq:hubbardI}
\mathbf{G}_{lat}^{-1} \approx
\left[\begin{matrix} \omega - \epsilon & - t_{\perp}\\
- t_{\perp} & \omega  - \epsilon \end{matrix}\right] -
\left[\begin{matrix}\Sigma_{11} & \Sigma_{12}\\\Sigma_{12} & \Sigma_{11}\end{matrix}\right]_{dimer}
\end{equation}

\begin{figure}\centering
\subfloat{\includegraphics[width=0.4\textwidth]{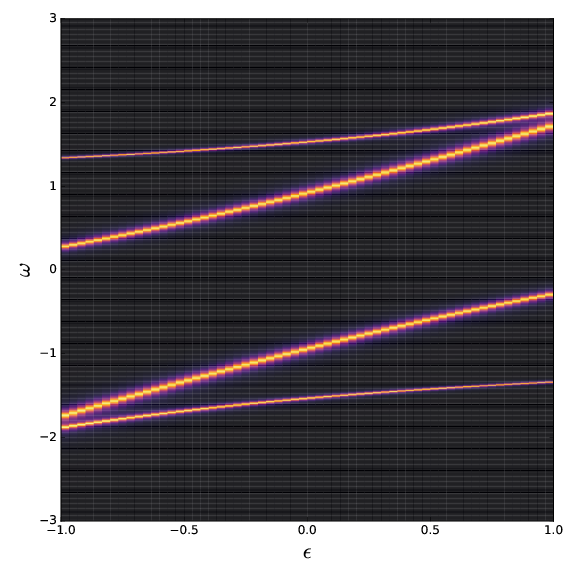}}
\subfloat{\includegraphics[width=0.4\textwidth]{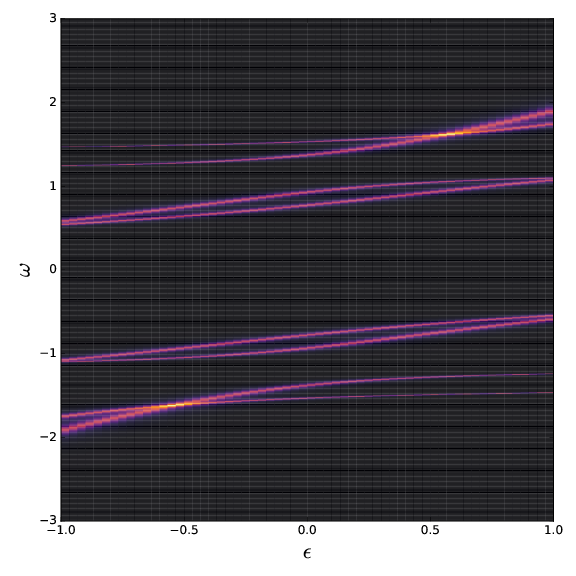}}
\caption{Lattice dispersion approximated with the isolated dimer Self-energy. Setup is $(U=2.15, t_\perp=0.3)$. Left panel at $\beta=100$ only ground state excitations, right panel $\beta=5$ presents excitations out of ground state and first excited states. \label{fig:dispersion_dimer}}
\end{figure}

The electronic structure in this limit is shown in figure
\ref{fig:dispersion_dimer} at 2 different temperatures. The
left panel is at inverse temperature \(\beta=100\), where we see two
highly dispersive bands along with two flatter ones. At high
temperature \(\beta=5\) more excitations are apparent as the first
excited state of the dimer becomes thermally populated. This
dramatically enriches the electronic structure. Interestingly, these
multiple excitations can be readily identified in the actual model
results shown in the Main Text.

\section{Optical Conductivity}
\label{sec:orgheadline10}

To calculate the optical conductivity one requires a geometrical
definition of the lattice. As a common practice, one turns to the
hypercube in infinite dimensions to keep on with the exact limit of
the DMFT approximation \cite{pruschke_hubbard_1993}. Using the Peierls
ansatz to find the current operator and aligning our dimer along the,
say, \(x\) direction in the hypercube one finds the current operator
along the \(x\) direction to be:

\begin{equation}
\label{eq:current_hc}
\hat{j}_x = \sum_{\sigma} e \left[ 2 {\text a} \frac{t}{\sqrt{2d}} \sin{(k_x {\text a})}
\left(\hat{b}^\dagger_{k_x,\sigma}\hat{b}_{k_x,\sigma} + \hat{a}^\dagger_{k_x\sigma}\hat{a}_{k_x,\sigma} \right) +
i \eta t_\perp \left(\hat{b}^\dagger_{k_x,\sigma}\hat{a}_{k_x,\sigma} - \hat{a}^\dagger_{k_x,\sigma}\hat{b}_{k_x,\sigma} \right)\right]
\end{equation}

where \(e\) is the electron charge, \(i\) is the imaginary unit, a is
the lattice unit cell length, and \(\eta \in (0, {\text a})\) is the separation
between atoms of the dimer. When one diagonalizes the lattice
Hamiltonian one can see it in terms of quasiparticles that form a
bonding \((\hat{b})\) and an anti-bonding \((\hat{a})\) band, one uses then the
operators \(\hat{b}^\dagger_{k_x,\sigma},(\hat{b}_{k_x,\sigma})\) to create
(annihilate) quasiparticles in the bonding band with momentum \(k_x\)
and spin \(\sigma\) and analogously the operators
\(\hat{a}^\dagger_{k_x,\sigma}, (\hat{a}_{k_x,\sigma})\) for the anti-bonding
band. In infinite dimensions to keep the kinetic energy constant one
has to scale the hopping amplitude \(t\rightarrow \frac{t}{\sqrt{2d}}\)
\cite{Georges1996}. Then following the procedure established in
\cite{pruschke_hubbard_1993} we arrive to the expression

\begin{align}
\nonumber
\bar{\sigma}_{xx}(\omega) &=\frac{2 \pi e^2 {\text a}^2 t^2}{d} \int d\omega' \frac{f(\omega') -f(\omega'+\omega)}{\omega}\times\\
\nonumber
&\int dE \rho(E)\left[ A_a(E,\omega) A_a(E,\omega+\omega') + A_b(E,\omega) A_b(E,\omega+\omega') \right] \\
\nonumber
&+2 \pi e^2 \eta^2 t_\perp^2 \int d\omega' \frac{f(\omega') -f(\omega'+\omega)}{\omega}\times\\
& \int dE \rho(E) \left[A_b(E,\omega)A_a(E,\omega+\omega') + A_a(E,\omega)A_b(E,\omega+\omega')\right]
\label{eq:Optcond_gf_E_sum_lD}
\end{align}

\begin{figure}[htb]
\centering
\includegraphics[width=.9\linewidth]{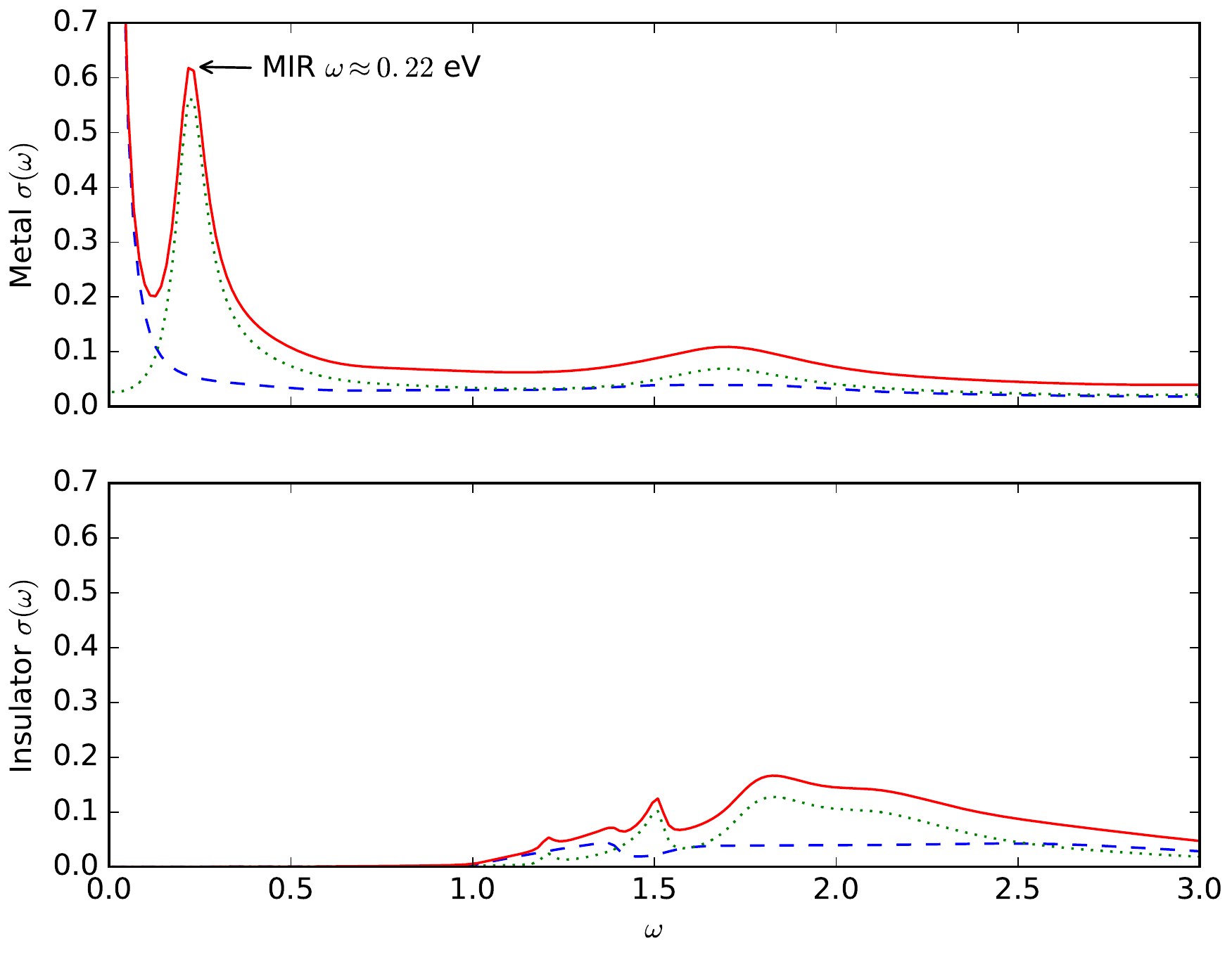}
\caption{Decomposition of Optical conductivity contributions for metal and insulator. Blue dashed lines are the intraband response, dotted green is the interband excitations, and red is the sum of the two. In the metal there is a MIR peak from interband transitions(cf. figure \ref{fig:IPT_ab_spectra_character})  \label{fig:decomposed_optcond}}
\end{figure}

Where \(\rho(E)=\frac{\exp(-E^2/(2t^2))}{\sqrt{2\pi t^2}}\) is the
density of states of the hypercube, and \(A_b, A_a\) are the spectral
functions of the bonding and anti-bonding bands.
From the previous equation we find that, in general, both intra-band
and inter-band transitions are present. Notice that the latter are usually disregarded
(e.g. Ref. \cite{Fuhrmann2006}). The two contributions are weighted with different
factors that depend on the specific geometry. For the sake of simplicity, we set those geometric
prefactors equal to unity, as they are expected to be of the same order ( a \(\sim \eta\),
\(t\sim t_\perp\)  and \(d=3\) ). In figure
\ref{fig:decomposed_optcond}, we plot for the metal and the insulator
case the individual contribution of the interband transitions from the
bonding to antibonding bands(refer to figure \ref{fig:IPT_ab_spectra_character}) and of the intraband excitations.

\putbib[biblio]
\end{bibunit}

\end{document}